\documentclass[twocolumn]{aa} 
\usepackage{graphicx}
\usepackage{nameref}
\usepackage{bm}
\usepackage[varg]{txfonts}

\begin{document} 
	
	\title{Minimal Roles of Solar Subsurface Meridional Flow in the distributed-shear Babcock-Leighton Dynamo}

	\author{Jie Jiang\inst{1}\fnmsep\inst{2}
		 \and		
		Zebin Zhang\inst{1}\fnmsep\inst{2}
	}
	
	\institute{School of Space and Earth Sciences, Beihang University, Beijing, China\\
		\email{jiejiang@buaa.edu.cn}
		\and
		Key Laboratory of Space Environment Monitoring and Information Processing of MIIT, Beijing, China		
	}
	\titlerunning{Minimal Roles of Subsurface Meridional Flow on the Solar Dynamo}
	
	\date{Received xx, 2025; accepted xx, 2025}

	\abstract
	{The subsurface meridional flow has long been recognized as a critical factor in driving the solar cycle. Specifically, the equatorward return flow in the tachocline is widely believed to be responsible for the formation of the sunspot butterfly diagram and determine the solar cycle period within the framework of flux transport dynamo (FTD) models.}
	{We aim to investigate whether the subsurface meridional flow also plays a significant role in the recently developed distributed-shear Babcock-Leighton (BL) dynamo model, which operates within the convection zone, rather than the tachocline.}	
	{Various meridional flow configurations, including a deep single cell, a shallow single cell, and double cells, are applied in the distributed-shear BL dynamo model to explore the mechanisms driving the butterfly diagram and variations in the cycle period.}
	{Subsurface meridional flow plays a minimal role in the distributed-shear BL dynamo. A solar-like butterfly diagram can be generated even with a double-cell meridional flow. The diagram arises from the time- and latitude-dependent regeneration of the toroidal field, governed by latitude-dependent latitudinal differential rotation and the evolution of surface magnetic fields. The cycle period is determined by the surface flux source and transport process responsible for polar field generation, which corresponds to the $\alpha$-effect in the BL-type dynamo. The cycle period may exhibit varying dependence on the amplitude of the subsurface flow.}
	{The distributed-shear BL dynamo differs fundamentally from the FTD models, as it does not rely on the subsurface flux transport. This distinction aligns the distributed-shear BL dynamo more closely with the original BL dynamo and the conventional $\alpha\Omega$ dynamo. Although the subsurface meridional flow plays a negligible role in our distributed-shear BL dynamo, the poleward surface flow is essential.}
	
	\maketitle
	
\section{Introduction}\label{intro}
The Sun exhibits cyclical magnetic activity \citep{Hathaway2011}, characterized by the quasi-11-year sunspot cycle and the quasi-22-year magnetic cycle \citep{Schwabe1849,Hale1919}. Beyond the variation in sunspot numbers, there is a systematic pattern in their latitudinal emergence, known as the sunspot butterfly diagram \citep{Maunder1904}. This diagram reveals a trend in which the latitude of sunspot emergence shifts progressively closer to the equator at an average velocity of approximately 2$^\circ$ - 3$^\circ$ per year as the cycle evolves \citep{Jiang2011a, Cameron2016, Biswas2022}. Understanding the mechanisms that govern the butterfly diagram and the factors that modulate the 11-year cycle period is crucial for gaining insight into the solar magnetic cycle \citep{Cameron2023}.

The solar magnetic cycle is driven by a dynamo process, which sustains cyclical evolution of toroidal and poloidal magnetic fields through turbulent convection and differential rotation, converting kinetic energy from plasma motions into magnetic energy \citep{Charbonneau2020}. Sunspots are believed to form due to the buoyancy of subsurface toroidal flux \citep{Parker1955b,Fan2021}. Consequently, the butterfly diagram of sunspots reflects the equatorward migration of subsurface toroidal fields, and the sunspot number serves as an indicator of the toroidal flux generated within the convection zone. Together, the butterfly diagram and the quasi-cyclic evolution of sunspot numbers provide key insights into the dynamics of the solar magnetic cycle, while also offering critical constraints for dynamo models.

In the 1990s, the flux transport dynamo (FTD) models of Babcock-Leighton (BL) type were developed \citep{Wang1991,Durney1995,Choudhuri1995}. From this point onward, we refer to them as the FTD models. In contrast to conventional mean-field $\alpha\Omega$ dynamo models \citep{Parker1955a, Steenbeck1969}, which involved two key processes (the regeneration of toroidal and poloidal fields from each other), FTD models involve an additional process \citep{Karak2014a,Choudhuri2023}. The three processes were specified as follows: (i) The toroidal field was produced from the poloidal field in the tachocline due to strong radial shear, corresponding to the $\Omega$-effect. (ii) The poloidal field was produced from the toroidal field by the BL mechanism at the surface, corresponding to the $\alpha$-effect. (iii) The poloidal and toroidal fields were connected through subsurface meridional circulation, which advected the poloidal field down to the tachocline, along with turbulent diffusion. In principle, the BL mechanism describes the poloidal field regeneration, where the toroidal field gives rise to sunspots with tilt angles \citep{Hale1919} due to magnetic buoyancy, which then decay on the surface \citep{Babcock1961,Leighton1969}. In practice, the BL mechanism is often treated with crude approximation methods in the FTD models \citep{Choudhuri2016}. 

In the FTD models, the observed poleward flow at the solar surface is assumed to have an equatorward return flow typically about 2 m s$^{-1}$ at the bottom of the convection zone \citep{Dikpati1999, Chatterjee2004}. The subsurface flow played a dominant role in driving the butterfly diagram and modulating solar cycle variations. The equatorward return flow can overpower the poleward dynamo wave predicted by the Parker–Yoshimura sign rule \citep{Parker1955a, Yoshimura1975} and forced the toroidal field to move equatorward, which led to butterfly diagrams \citep{Choudhuri1995, Hazra2014}. Moreover, the subsurface flow connected the BL source term at the surface with the $\Omega$-effect at the base of the convection zone. The faster the poloidal magnetic flux generated at the surface was transported to the bottom, the sooner the radial differential rotation can act on it to generate toroidal fields. This rapid transport accelerated the dynamo process, leading to an earlier onset of the next solar cycle. During the past few decades, the FTD modes have been the primary tool for understanding the solar cycle, from modeling both its regularities and irregularities \citep[e.g.,][]{Dikpati1999, Chatterjee2004, Karak2010,Kitchatinov2018,Lemerle2017} to making predictions \citep[e.g.,][]{Dikpati2006, Choudhuri2007,Jiang2007, Guo2021}. For reviews, see \cite{Bhowmik2023, Jiang2023, Karak2023}.
 
However, recent developments have raised questions about whether the FTD model is the correct framework for the solar cycle, as discussed below. One of the primary prerequisites for FTD models is that the return flow at the bottom of the convection zone must be equatorward. Without this equatorward flow, the solar-like butterfly diagram, which reflects the equatorward migration of the toroidal field, cannot be generated \citep{Hazra2014}. For decades, observations have confirmed the presence of a poleward surface flow on the Sun \citep[e.g.,][]{Ward1973,Makarov1983}. To conserve mass, this surface flow necessitates an equatorward return flow within the solar interior. However, helioseismic studies indicate a controversial profiles of the subsurface meridional flow. \cite{Zhao2013, Chen2017} identify double cells in depth: a poleward flow from $R_\odot$ to about 0.91 $R_\odot$, an equatorward flow between 0.82-0.91 $R_\odot$, and poleward flow again below 0.82 $R_\odot$. \cite{Schad2013} report multiple cells in both depth and latitude. \citet{Jackiewicz2015} and \citet{Rajaguru2015, Gizon2020} infer a single-cell circulation with return flows starting in shallower and deeper regions, respectively. Moreover, the precise configuration of this flow is challenging to determine from first principles, as it arises from a subtle imbalance between two dominant terms in the governing equations \citep{Kitchatinov2005, Kitchatinov2013}. Convective MHD simulations  have not yet reached a consensus either.  \cite{Featherstone2015,Hotta2022} show that the solar-like fast equator leads to a poleward flow around the base of the convection zone and a equatorward flow in the middle of the convection zone. \cite{Brun2017a,Guerrero2019} suggest that solar-like models tend to have multicellular flow structures. \cite{Karak2015} find that the simulated meridional flow for anti-solar rotations is consistent with the usual assumption in FTD models. For solar-like rotations, however, the meridional flow weakens and develops multiple cells, although some equatorward flow near the base of the convection zone is found in their simulated multicellular flow structures. In summary, both helioseismic inversions and MHD simulations indicate that the equatorward return meridional flow at the base of the convection zone is not robustly established, which poses a challenge to the FTD models. 

Besides the aforementioned potentially problematic prerequisite, FTD models also rely on a key assumption that might also meet challenges. That is, the toroidal fields are generated at the tachocline, a thin layer at the bottom of the convection zone. The tachocline's strong radial shear and subadiabatic stratification may effectively amplify and store the magnetic fields before being subjected to magnetic buoyancy instabilities. The scenario was supported by thin flux tube simulations \citep[for reviews, see][]{Fan2021}, which solidified the prevailing view that tachocline was the location of the toroidal field generation. However, recent MHD simulations indicate that solar-like large-scale magnetic fields can be produced entirely within a convection zone \citep[e.g.,][]{Fan2014, Kapyla2017}. Rope-like structures of magnetic flux can spontaneously form and rise in a solar-like manner without the need for a tachocline \citep{Nelson2014, Chen2022}. Similarities in large-scale field generation between partially and fully convective stars have also been presented in simulations \citep[e.g.,][]{Kapyla2021}. Furthermore, stellar activity indices, such as flare and X-ray luminosity, show that fully convective M-type stars follow the same activity-rotation relationship as partially convective stars \citep{Wright2016, Yang2019}. These advancements suggest that the assumption of the tachocline as the location of the toroidal field generation could be questionable. 

To address these challenges, \citet{Zhang2022} recently develop a new BL - type dynamo model that operates throughout the bulk of convection zone. The radial outer boundary condition and near surface pumping enable the surface field evolution in the model consistent with that from semi-empirical surface flux transport (SFT) models \citep{Jiang2014, Yeates2023}. As a result, the model reproduces a large-scale dipolar field connecting the two poles at cycle minimum, consistent with what observations indicate. The latitudinal shear in the bulk of the convection zone acts on the large-scale field to regenerate the toroidal fields there. As a result, the tachocline plays a negligible role in this model. Since the generation of the toroidal field is distributed across the entire convection zone, rather than being confined to the thin layer of tachocline, we henceforth refer to this model as the distributed-shear BL dynamo. 

This work aims to investigate the effects of subsurface meridional flow on the distributed-shear BL dynamo, comparing it with FTD models. The paper is organized as follows. Section \ref{sec:model} outlines the distributed-shear dynamo model used in this study. Section \ref{subsec:ImactButterfly} examines the impact of different meridional flow profiles on the migration of the toroidal magnetic field. In Sect. \ref{subsec:ImactCyclePeriod}, we explore how the speed of the meridional flow influences the cycle period. Finally, Sect. \ref{sec:conclustion} provides a summary of the results and a discussion of their implications.

\section{The distributed-shear Babcock-Leighton dynamo model}\label{sec:model}
The distributed-shear BL dynamo model describes the evolution of the axisymmetric large-scale toroidal field, $B_\phi(r,\theta,t) \hat{\bm{e}}_\phi$, and the poloidal field, $\bm{B}_{p}$ = $\nabla \times A(r,\theta,t)\hat{\bm{e}}_\phi$, where $A(r,\theta,t)$ represents the magnetic vector potential in the $\phi$ direction. The dynamo equations governing the evolution of $B_\phi(r,\theta,t)$ and $A(r,\theta,t)$ are expressed as:
\begin{equation} \label{eq1:dynamo_A}
	\frac{\partial A}{\partial t}+\frac{1}{s}[(\bm{u}_{p}+\gamma_{r}\hat{\bm{e}}_{r})\cdot\nabla](sA)
	=\eta\left(\nabla^{2}-\frac{1}{s^{2}}\right)A+S_{BL}, 
\end{equation}

\begin{equation}  \label{eq2:dynamo_B}
	\begin{split}
		\frac{\partial B_\phi}{\partial t}+\frac{1}{r}\left[\frac{\partial(u_{r}+\gamma_{r})rB_\phi}
		{\partial r}+\frac{\partial(u_{\theta}B_\phi)}{\partial\theta}\right]=\eta\left(\nabla^{2}-\frac{1}
		{s^2}\right)B_\phi+ \\
		s(\bm{B}_{p}\cdot\nabla\Omega)+\frac{1}{r}\frac{d\eta}
		{dr}\frac{\partial(rB_\phi)}{\partial r}, 
	\end{split}
\end{equation}  
where $s=r\sin\theta$. The symbols $\Omega(r,\theta)$ and $\bm{u}_p(r,\theta)$ = $u_r(r,\theta) \hat{\bm{e}}_r$ + $u_\theta(r,\theta) \hat{\bm{e}}_\theta$ represent the differential rotation and meridional flow, respectively, and $\gamma_{r}$ and $\eta$ represent the turbulent radial pumping and the turbulent diffusivity, respectively.

\begin{figure*}[!htp]
	\centering
	\includegraphics[width=16cm]{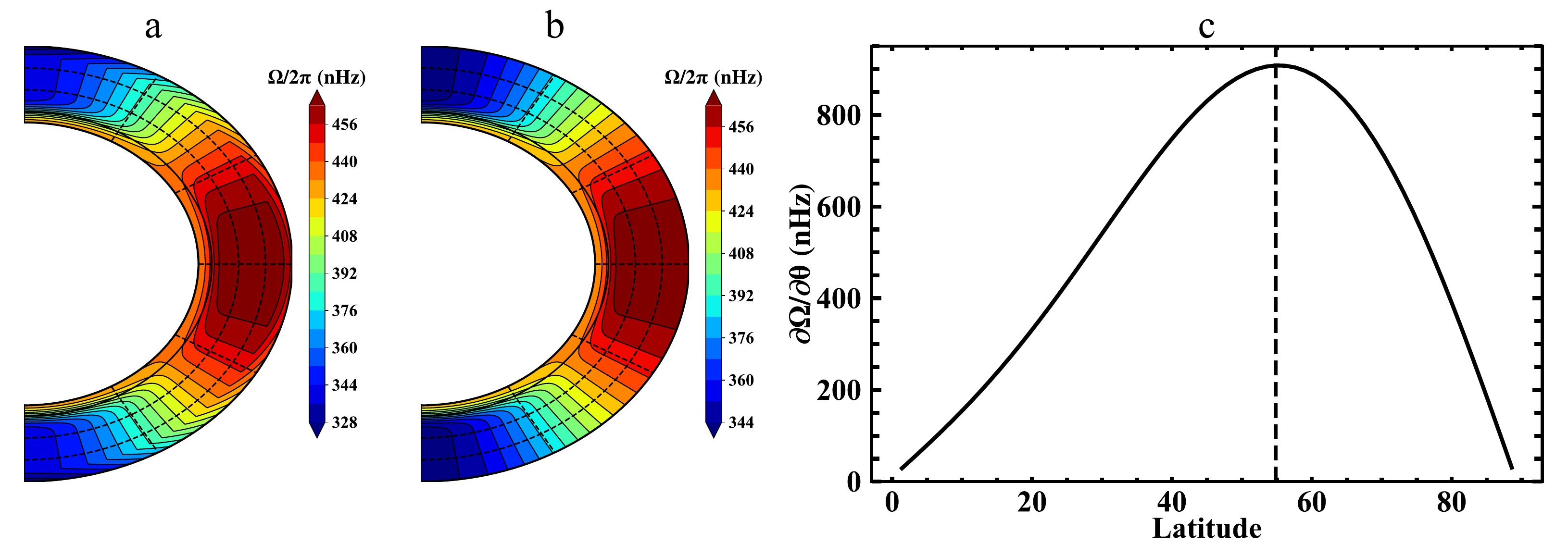}	
	\caption{Differential rotations used in the paper. (a) Contours of differential rotation $\Omega(r, \theta)$ used in Sect.\ref{subsec:ImactCyclePeriod}. (b) Contours of the rotation without the near-surface shear layer used in Sect.\ref{subsec:ImactButterfly}. (c) Latitude dependence of the latitudinal differential rotation $\partial \Omega/ \partial \theta$ at the surface.}
	\label{Figure2_DF}
\end{figure*}

Figure \ref{Figure2_DF}(a) displays the differential rotation profile adopted in the model by \citet{Zhang2022}. To eliminate the influence of the dynamo wave in the near-surface shear layer \citep{Pipin2011,Karak2016}, the simulations in Sect. \ref{subsec:ImactButterfly} adopt the modified differential rotation profile shown in Fig. \ref{Figure2_DF} (b), where the radial gradient near the surface has been removed. Figure \ref{Figure2_DF}(c) illustrates $\partial \Omega/ \partial \theta$ as a function of latitude at the surface. The latitudinal gradient of differential rotation peaks around $\pm$55$^\circ$ latitudes and decreases at both higher and lower latitudes. The property closely aligns with helioseismic results \citep{Schou1998}. 

The pumping profile used in this study is identical to that in Eq. (13) of \citet{Zhang2022}, except with a strength of $\gamma_{0}$ = 25 ms$^{-1}$ and a penetration depth of $r_p = 0.9 R_\odot$. Strong radial pumping is required to align the surface behavior of the dynamo model with SFT models and observations \citep{Cameron2012}. The pumping effect suppresses cross-surface diffusion of the magnetic field and ensures that the poloidal field remains predominantly radial near the surface. This ultimately results in the formation of a large-scale dipolar field \citep{Cameron2012, Jiang2013, Karak2016, Karak2017}.

As will be demonstrated in Sects. \ref{subsec:ImactButterfly} and \ref{subsec:ImactCyclePeriod}, the poloidal source term $S_{BL}$ in Eq. (\ref{eq1:dynamo_A}) plays a critical role in the operation of the distributed-shear BL-type dynamo. Its significance in stellar magnetic cycle has been highlighted by \citet{Zhang2024}. The term is defined as 
\begin{equation}
	\label{eq:S_BL}
	S_{BL}=\alpha(r, \theta)\overline{B_\phi},
\end{equation}
where $\overline{B_\phi}$ is the toroidal field averaged over the whole convection zone, following the same form as Eq.(8) of \citet{Zhang2022}. The form was first introduced by \cite{Karak2016}. The $\alpha$-term is confined to the near-surface region and expressed as 
\begin{equation}
	\alpha(r,\theta)=\frac{\alpha_{0}f(\theta)}{2}\left[1+\rm
	erf\left(\frac{r-0.95R_\odot}{0.01R_\odot}\right)\right],\label{eq:alpha}
\end{equation}
where $\alpha_{0}$ determines the amplitude of the $\alpha$-term, and $f(\theta)$ describes its latitudinal dependence, given by
\begin{equation}
	f(\theta)=\cos\theta \sin^n\theta,\label{eq:alpha_f}
\end{equation}
where $n$ is a free parameter that constrains the latitudinal location of poloidal fields. In Sect. \ref{subsec:ImactButterfly}, $n$ is set to $n=$10, while in Sect. \ref{subsec:ImactCyclePeriod}, $n$ is set to $n=$6. The smaller $n$ value corresponds to slightly higher-latitude concentration of the source term. The formulation of the $S_{BL}$ term differs from the reference model of \citet{Zhang2022} in one key aspect. That is, we do not include the nonlinear algebraic quenching term here. Instead, this work focuses on studying the dynamo process in linear critical regimes, as different amplitude-limiting mechanisms can influence dynamo behavior in diverse ways. Understanding the linear dynamo process is crucial before exploring the effects of nonlinear amplitude-limiting mechanisms in our model. All simulations in this work are conducted in the linear critical regime, where $\alpha_{0}$ is set to its critical value $\alpha_{c}$, and the critical value is determined when the growth rate of the dynamo solution approaches zero. 

\begin{figure*}[!htp]
	\centering
	\includegraphics[width=16cm]{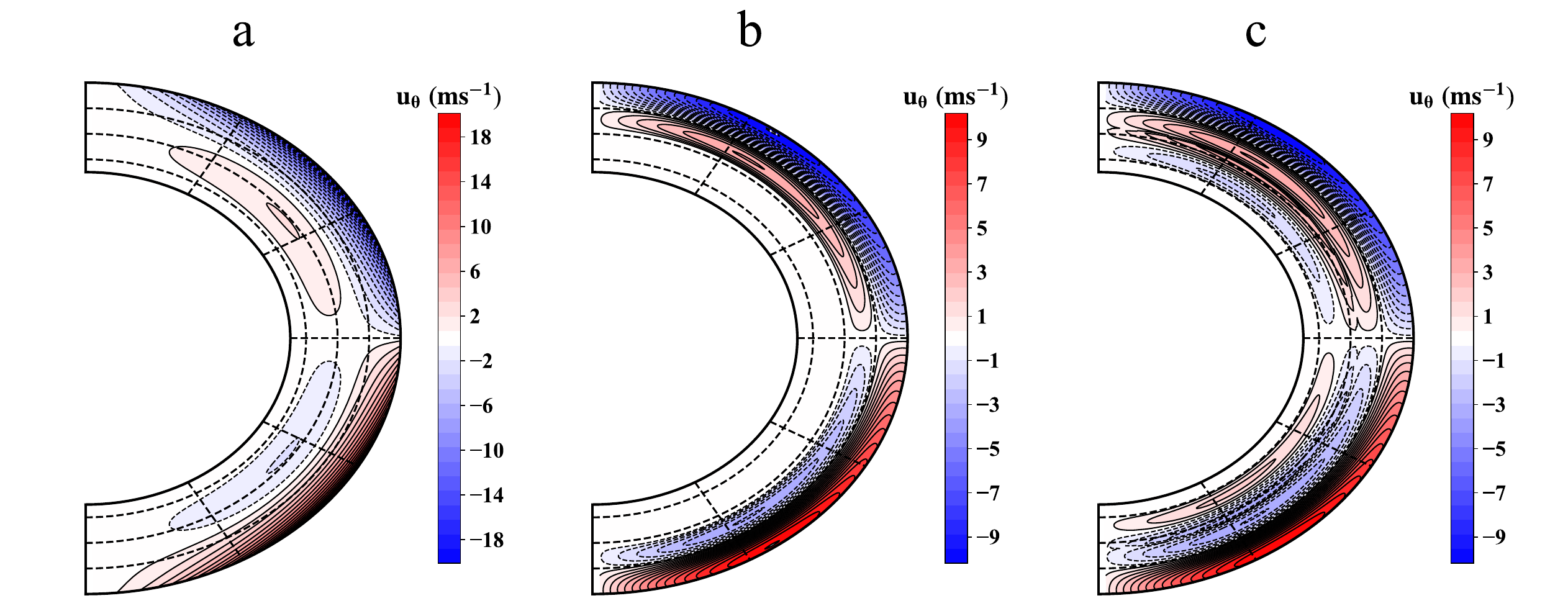}	
	\caption{Three profiles of the latitudinal meridional flow $u_\theta$ used in the paper: (a) a deep one-cell flow (MF1), (b) a shallow one-cell flow (MF2), and (c) a double-cell flow (MF3). Positive (negative) velocities indicate equatorward (poleward) flow directions.}
	\label{Figure1_MF}
\end{figure*}

In this work, we consider three representative profiles of meridional flow, as illustrated in Fig. \ref{Figure1_MF}: a single-cell flow with deep return flow (MF1), a single-cell flow with shallow return flow (MF2), and a double-cell flow (MF3). The meridional flow is defined in terms of stream function $\Psi$, which is given by $\rho \bm{u}_p = \nabla \times \Psi \hat{\bm{e}}_\phi$ with the density profile $\rho = C(R_\odot/r-0.95)^{3/2}$. The stream function for the deep penetration of a single-cell flow MF1 is $\Psi_{1}$, defined as
\begin{align}
	\Psi_{1} r \sin\theta =& \psi_0(r-R_p)\sin\left[\frac{\pi(r-R_p)}{R_\odot-R_p}\right] \notag \\
	&(1-e^{-\beta_1\theta^\varepsilon })[1-e^{\beta_2(\theta-\pi/2)}]e^{-[(r-r_0)/\Gamma]^2 }\sin\theta,\label{eq3}
\end{align}
where $\beta_1 = 1.5,\ \beta_2 = 1.3,\ \varepsilon = 2.0000001,\ r_0 = (R_\odot-R_b)/3.5,\ \Gamma =3.47 \times 10^8$ m and $R_p = 0.7R_\odot$. The value of $\psi_0/C$ determines the amplitude of the flow. For the study in Sect. \ref{subsec:ImactButterfly}, the surface velocity is set to reach a maximum of 20 m~s$^{-1}$ at $\pm45^\circ$. This velocity profile is the same as that used in \cite{Zhang2022} and is widely adopted in FTD models. The values will be adjusted in Sect. \ref{subsec:ImactCyclePeriod}, as stated there. To facilitate comparison with the results from \cite{Hazra2014} for the same flow configuration, we adopt their definitions for the stream functions of $\text{MF2}$ and $\text{MF3}$. The stream function $\Psi_2$ for $\text{MF2}$ is defined by $\Psi_2$ = $\Psi_u$, which corresponds to a single cell concentrated in the upper part of the convection zone. The function $\Psi_u$ is shown by
\begin{align}\label{MF}
	\Psi_u = & \, \psi_{0} \left[ 1 - \text{erf}\left(\frac{r - R_{s}}{s_0}\right)\right](r - R_{p}) \notag \\
	& \times \sin\left[\frac{\pi(r - R_{p})}{s_1 (R_{\odot} - R_{p})}\right]\left\{1 - e^{-\beta_1 \theta ^{\epsilon}}\right\} \notag \\
	& \times \left\{1 - e^{\beta_2 (\theta - \pi/2)}\right\} e^{-\left(\frac{(r - r_0)}{\Gamma}\right)^2}.
\end{align}
The parameters in $\Psi_u$ are specified as follows: $\beta_1 = 3.5$, $\beta_2 = 3.3$, $s_0 = 1.0$, $s_1 = 1.0$, $r_0 = (R_{\odot} - R_b)/3.5$, $\Gamma = 3.4 \times 10^8 \, \text{m}$, $R_{p} = 0.815 \, R_{\odot}$, and $R_{s} = 0.91 \, R_{\odot}$. The value of $\psi_0 / C$ is set to 16.1 such that the surface flow reaches a maximum value of $10\ \text{m s}^{-1}$. The single-cell flow just penetrations to $R_{p}=0.815 \, R_{\odot}$, with the return flow starting at $0.9 \, R_{\odot}$. The configuration of $\text{MF2}$ is presented in Fig. \ref{Figure1_MF}(b).  The $\text{MF3}$ corresponds to two radially stacked cells and its stream function is defined as $\Psi_3$ = $\Psi_u$ + $\Psi_l$, where $\Psi_l$ corresponds to the cell in the low part of the convection zone. It is derived from the same function as $\Psi_u$ but with different parameter sets. For $\Psi_l$, the parameters are defined as follows: $\beta_1 = 3.2$, $\beta_2 = 3.0$, $s_0 = 1.8$, $s_1 = 0.825$, $r_0 = (R_{\odot} - R_b)/3.5$, $\Gamma = 3.24 \times 10^8 \, \text{m}$, $R_p = 0.65 \, R_{\odot}$, and $R_{s} = 0.825 \, R_{\odot}$. The value of $\psi_0 / C$ is set to 0.45, resulting in maximum flow amplitudes of 3.46 m s$^{-1}$ for equatorward part and 1.86 m s$^{-1}$ for poleward part. The flow described by $\Psi_l$ forms the single counter-cell at the base of the convection zone. Together, $\Psi_l$ and $\Psi_u$ define the double-cell flow as illustrated in Fig. \ref{Figure1_MF}(c), which contains a poleward flow within the range of $0.7$ to $0.76 \, R_{\odot}$. 

The turbulent diffusivity $\eta$ in Eqs.(\ref{eq1:dynamo_A}) and (\ref{eq2:dynamo_B}) adopts the same radial dependence as Eq. (14) of \cite{Zhang2022}. The surface diffusivity, $\eta_{S}$, and the diffusivity in the radiative zone, $\eta_{RZ}$, are set to the same values as in that study. But the diffusivity in the bulk of the convection zone, $\eta_{cz}$, is variable. In Sect. \ref{subsec:ImactButterfly}, it is adjusted to ensure a cycle period of 11 years. The specific values of the diffusivity are set as follows: $\eta_{cz} = 3.7 \times 10^{11}$ cm$^{2}$ s$^{-1}$ for MF1, $\eta_{cz} = 1.1 \times 10^{11}$ cm$^{2}$ s$^{-1}$ for MF2, and $\eta_{cz} = 1.6 \times 10^{11}$ cm$^{2}$ s$^{-1}$ for MF3. In Sect. \ref{subsec:ImactCyclePeriod}, in order to ensure the dynamo successfully operates when maximum flow speed ranging from 10 ms$^{-1}$ to 20 ms$^{-1}$, the specific values of $\eta_{cz}$ are set as follows: $\eta_{cz} = 6.0 \times 10^{11}$ cm$^{2}$ s$^{-1}$ for MF1, $\eta_{cz} = 1.0 \times 10^{11}$ cm$^{2}$ s$^{-1}$ for MF2, and $\eta_{cz} = 1.0 \times 10^{11}$ cm$^{2}$ s$^{-1}$ for MF3.  

The boundary conditions are identical to those in \cite{Zhang2022}. Given that the solar magnetic field is predominantly anti-symmetric, we adopt a dipolar parity configuration as the initial condition. Specifically, the initial toroidal field is defined as
\begin{equation}
	B_\phi(r, \theta)|_{t=0} = \sin(2\theta)\sin[\pi(r-0.72R_\odot)/0.28R_\odot],
	\label{eq10}
\end{equation}
and the poloidal field is initialized to zero. 

Since the dynamo models operate in the linear regime, the magnetic field strength can be arbitrarily scaled. To facilitate comparisons with observations, we scale the magnetic fields such that the maximum of the surface radial field is 10 G, consistent with the observed polar field strengths at solar cycle minimum \citep{Cameron2010, Jiang2011b, Munoz-Jaramillo2012, Wang2020}.

Our model is computed using a numerical code developed at Beihang University, which employs the Crank-Nicolson scheme combined with an approximate factorization technique \citep{Houwen2001}. The code has been rigorously validated against the dynamo benchmark established by \cite{Jouve2008}, ensuring its accuracy and reliability for the simulations.

\section{Results}\label{sec:results}
\subsection{Minimal Impact of Subsurface Meridional flow Profiles on the Migration of the Toroidal Magnetic Field} \label{subsec:ImactButterfly}
We first explore the impact of subsurface meridional flow on the equatorward migration of the toroidal field in the distributed-shear BL dynamo. The three profiles of subsurface meridional flow, as shown in Fig. \ref{Figure1_MF}, are considered as three cases. As noted in Sect. \ref{sec:model}, to exclude the effects of the equatorward dynamo wave produced by near surface radial shear layer, the differential rotation profile shown in Fig. \ref{Figure2_DF}(b) is used in this section.

\begin{figure*}[!htp]
	\centering
	\includegraphics[width=15cm]{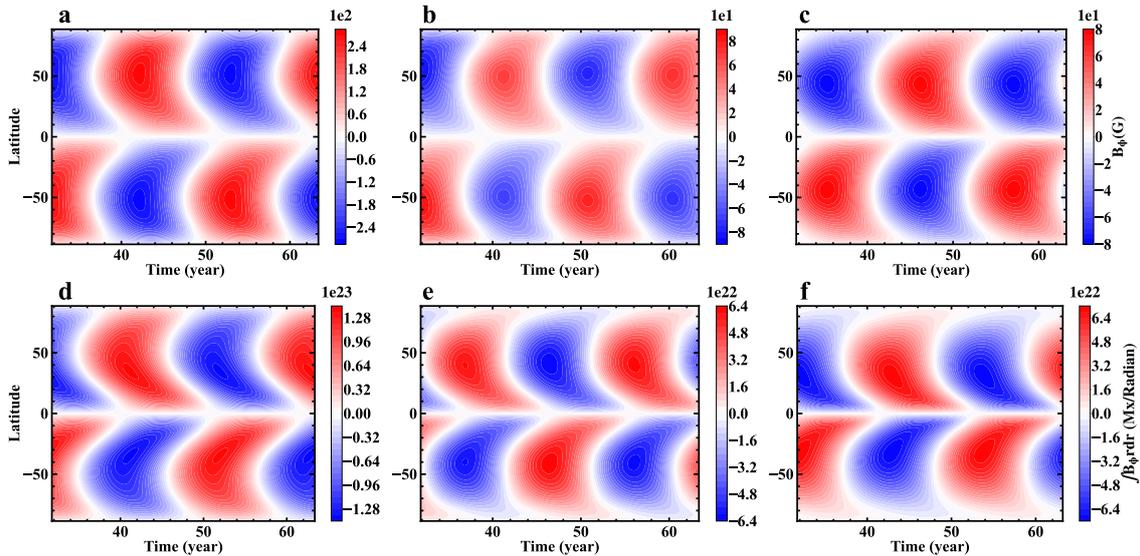}
	\caption{Time-latitude diagrams of the subsurface toroidal field in distributed-shear BL dynamo models. Upper and lower panels correspond to the toroidal field at the bottom of the convection zone $B_\phi(0.7R_\odot, \theta, t)$ and the radially integrated toroidal magnetic flux over $0.7R_\odot$ and $R_\odot$, $\Phi(0.7-1.0R_\odot, \theta, t)$, respectively. The left, middle, and right panels are the results obtained using the meridional flow profiles MF1, MF2, and MF3, respectively.}
	\label{Figure3_BphiThetaT}
\end{figure*}

In view that the tachocline was believed to the location of the sunspot-forming toroidal flux in previous dynamo models, we first investigate the time evolution of the toroidal field at the bottom of the convection zone $B_\phi(0.7R_\odot, \theta, t)$ for the 3 profiles of meridional flow. The results are shown in Figs. \ref{Figure3_BphiThetaT} (a), (b), and (c), which correspond to the dynamo models with a deep one-cell flow, a shallow one-cell flow, and a double-cell flow, respectively. In the framework of the FTD models, it is expected that only the deep penetration of the equatorward flow below $0.7R_\odot$ would result in the equatorward migration of the toroidal field below the $\pm55^\circ$ latitudes. However, the models incorporating equatorward return flows penetrating to $0.7R_\odot$ (Fig. \ref{Figure3_BphiThetaT}a), $0.815R_\odot$ (Fig. \ref{Figure3_BphiThetaT}b), or even poleward flows (Fig. \ref{Figure3_BphiThetaT}c) all produce similar migration patterns that differ strikingly from FTD model results. As noted by \citet{Hazra2014}, if there is either no flow or a poleward flow at the bottom of the convection zone, FTD models only reproduce poleward migration of the toroidal field at the low latitudes. In addition, in our models the toroidal field is produced and distributed in the convection zone. Hence, we also investigate the time evolution of the toroidal flux calculated by radially integrating the toroidal field over the range of 0.7-1.0$R_\odot$, $\Phi(0.7-1.0R_\odot, \theta, t)=\int_{0.7R_\odot}^{1.0R_\odot}B_\phi r dr$. The results are shown in the lower panels of Fig. \ref{Figure3_BphiThetaT}, which exhibit an even better equatorward migration pattern for each flow configuration. And the pattern remains almost unchanged when the integration range is adjusted. 

The well-defined equatorward migration pattern of the toroidal field in the absence of flow or even with a poleward flow at the bottom of the convection zone raises the important question about its generation mechanism. If this were due to the dynamo wave propagating along isorotation surfaces \citep{Yoshimura1975}, the migration patterns of $B_\phi(0.7R_\odot, \theta, t)$ should be poleward due to $\partial \Omega/ \partial r >$ 0 at the low latitudes of $0.7R_\odot$ and $\alpha > 0$ in the northern hemisphere. As have demonstrated by \citet{Zhang2022}, the radial shear in the tachocline actually plays a negligible role in our distributed-shear BL dynamo model. If we artificially remove the radial shear in the tachocline or set the bottom boundary at a depth of 0.725$R_\odot$, the dynamo behavior does not show significantly change. The generation of the toroidal field is dominated by the latitudinal shear in the bulk of the convection zone. Hence the migration patterns of $\Phi(0.7-1.0R_\odot, \theta, t)$ only relate with the latitudinal shear. And a purely latitudinal shear of the differential rotation should lead to radially propagating dynamo waves, and no latitudinal propagation wave. To identify the underlying mechanism responsible for the equatorward migration, we conduct the following test. 
 
\begin{figure*}[!htp]
	\centering
	\includegraphics[width=8cm]{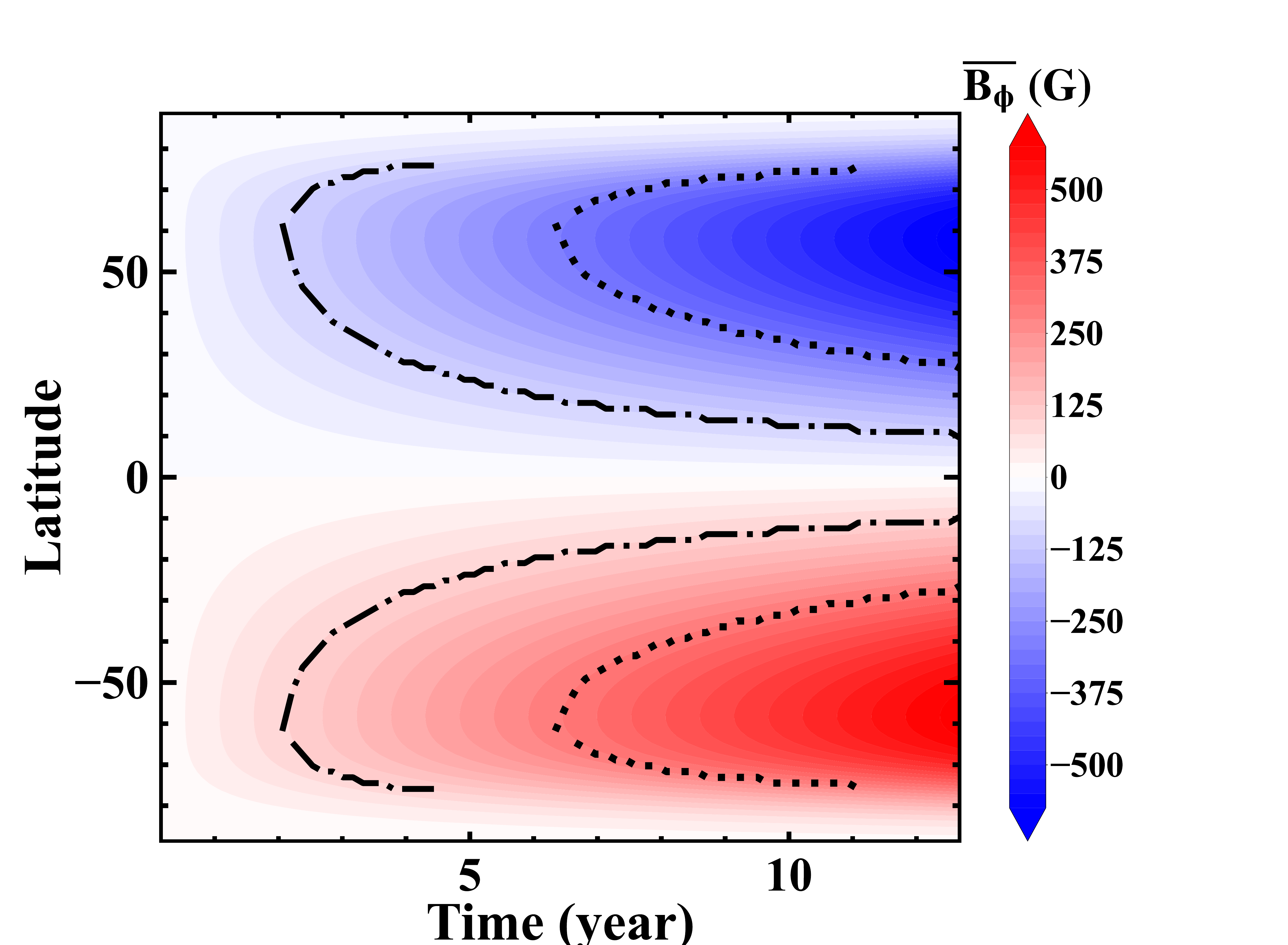}
	\includegraphics[width=7.5cm]{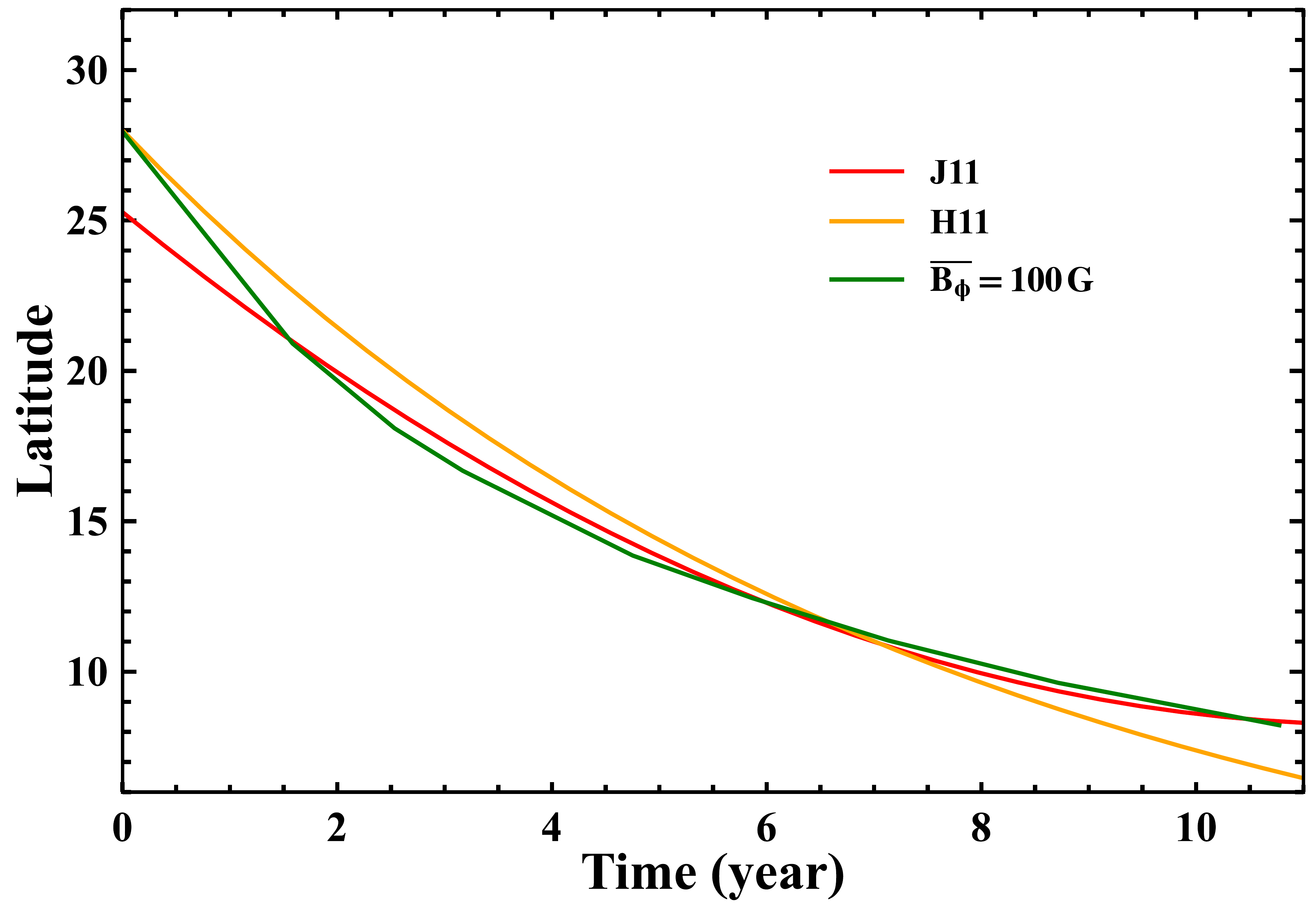}	
	\caption{A numerical test starting from an initial dipolar magnetic field with a maximum strength of 10 G to illustrate the role of the latitude-dependent latitudinal shear in generating the equatorward migration of the toroidal field. Left panel: regeneration pattern of the average toroidal magnetic field within the range 0.7$R_\odot$ to $R_\odot$, $\overline{B_\phi}$. The dot-dashed and dotted curves represent the time evolution of the latitudes at which $\overline{B_\phi}$ reaches 100 G and 300 G, respectively. Right panel: comparison of the time-latitude dependence of $\overline{B_\phi}$=100 G (green) with observations measured based on sunspot records by \citet[][H11, orange]{Hathaway2011} and by \citet[][J11, red]{Jiang2011a}.}
	\label{Figure4_LatDepShear}
\end{figure*}

The test begins with an initial dipolar magnetic field of maximum strength 10 G, ignoring diffusion and advection terms due to meridional flow as described in Eq. (\ref{eq2:dynamo_B}). Only the differential rotation in the form of Fig. \ref{Figure2_DF}b works on the dipolar field. Left panel of Fig. \ref{Figure4_LatDepShear} presents the regeneration pattern of the average toroidal magnetic field within the range 0.7$R_\odot$ to 1$R_\odot$, $\overline{B_\phi}$, resulting from latitude-dependent latitudinal shear. We may see that the toroidal field starts from zero and increases over time. It takes about $\Delta t =$ 2 years for $\overline{B_\phi}$ to reach 100 G (dot-dashed curve) around $\pm55^\circ$ latitudes, whereas it takes about $\Delta t =$ 10 years to reach the same strength at latitudes around $\pm10^{\circ}$. Right panel of Fig. \ref{Figure4_LatDepShear} compares the time-latitude dependence of $\overline{B_\phi}$=100 G with observations measured based on sunspot records \citep{Hathaway2011,Jiang2011a}, showing good agreement between the two.
For a stronger $\overline{B_\phi}$, i.e., $\overline{B_\phi}=$300 G (dotted curve), it takes longer time to reach the larger critical $\overline{B_\phi}$ as $\overline{B_\phi} \propto\partial \Omega/ \partial \theta \Delta t$. As shown in Fig.\ref{Figure2_DF}c, the latitudinal shear is latitudinal dependent, peaking around $\pm55^\circ$ latitudes and decreasing toward the poles and the equator. This leads to the rapid buildup of the toroidal flux around $\pm55^\circ$, with slower accumulation occurring at higher and lower latitudes. The time difference between the emergence at $\pm55^\circ$ latitudes and the equator mimics the butterfly diagram pattern. The latitude dependence of the latitudinal shear may thus play a dominant role in the equatorward migration pattern of the toroidal field presented in Fig. \ref{Figure3_BphiThetaT}. 
 
Beside the equatorward migration of the toroidal field at the activity belt, there is a poleward branch of the toroidal field above the $\pm55^\circ$ latitudes, as presented in Fig.\ref{Figure3_BphiThetaT} and left panel of Fig.\ref{Figure4_LatDepShear}. This poleward branch is demonstrated as an essential component for producing the poleward branches of the torsional oscillations originated from about the $\pm55^\circ$ latitudes \citep{Zhong2024}. Thus it provides support for the toroidal field generated in the bulk of the convection zone by the latitudinal shear. In addition, the test we conduct above is similar to Fig. 1 of \cite{Spruit2011} in essence.  

The mechanism for the latitudinal migration of the toroidal field was first proposed in one of the foundational papers of the BL dynamo, namely \cite{Babcock1961}. As such, this mechanism should be considered as the intrinsic and original explanation for the BL-type dynamo. It interprets the butterfly diagram as a regeneration pattern of toroidal fields, distinguishing it from later-developed FTD models. While FTD models are also classified as the BL type, they attribute the latitudinal migration of the toroidal field to transport processes driven by meridional flow. The mechanism for the latitudinal migration of the toroidal field raises another important  question related to the other seminal BL dynamo paper by \cite{Leighton1969}. The first case discussed in the Results Section of \cite{Leighton1969} examines a model with purely latitudinal differential rotation, which is similar to our case in 1D model. It produces oscillatory solutions with equatorward migration of the toroidal field, which are in close quantitative agreement with the solar butterfly diagram. The reason for the equatorward migration of the toroidal field once remained illusive. As point out by \cite{Cameron2017}, latitudinal migration does not violate the Parker-Yoshimura rule and is not excluded in the models for purely latitudinal differential rotation. But there was an error to deal with the regeneration term in the equation for the surface radial field. \cite{Cameron2017} note that, with a correct form of the regeneration term, the dynamo solutions decay. Why can our model produce the stable oscillatory solutions?

\begin{figure*}[!htp]
	\centering
	\includegraphics[width=18cm]{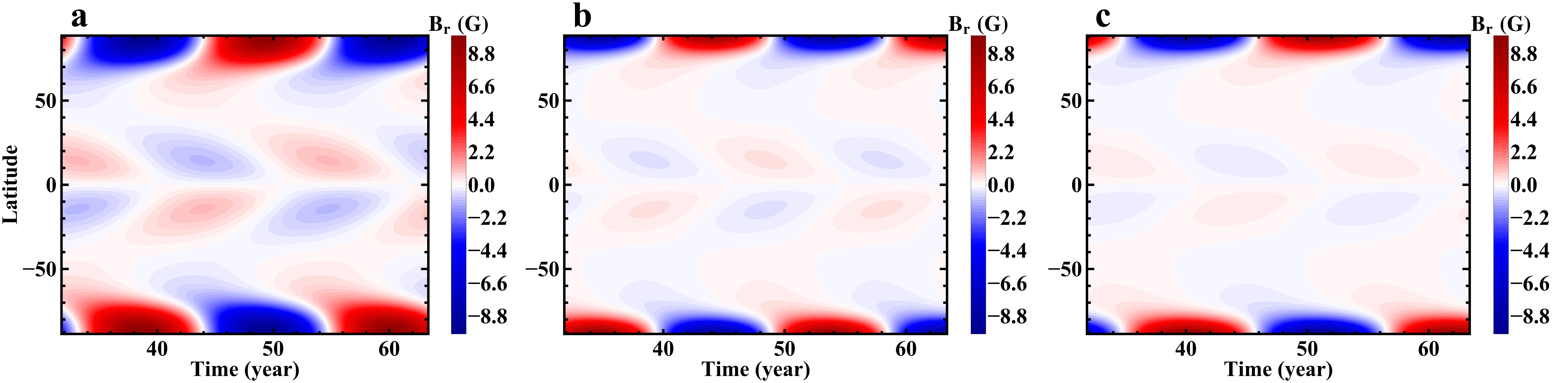}	
	\caption{Time-latitude diagrams of the surface radial fields in distributed-shear BL dynamo models. The left, middle, and right panels show the results obtained using the meridional flow profiles MF1, MF2, and MF3, respectively.}
	\label{Figure5_BrThetaT}
\end{figure*}

\begin{figure*}[h]
	\centering
	\includegraphics[width=15cm]{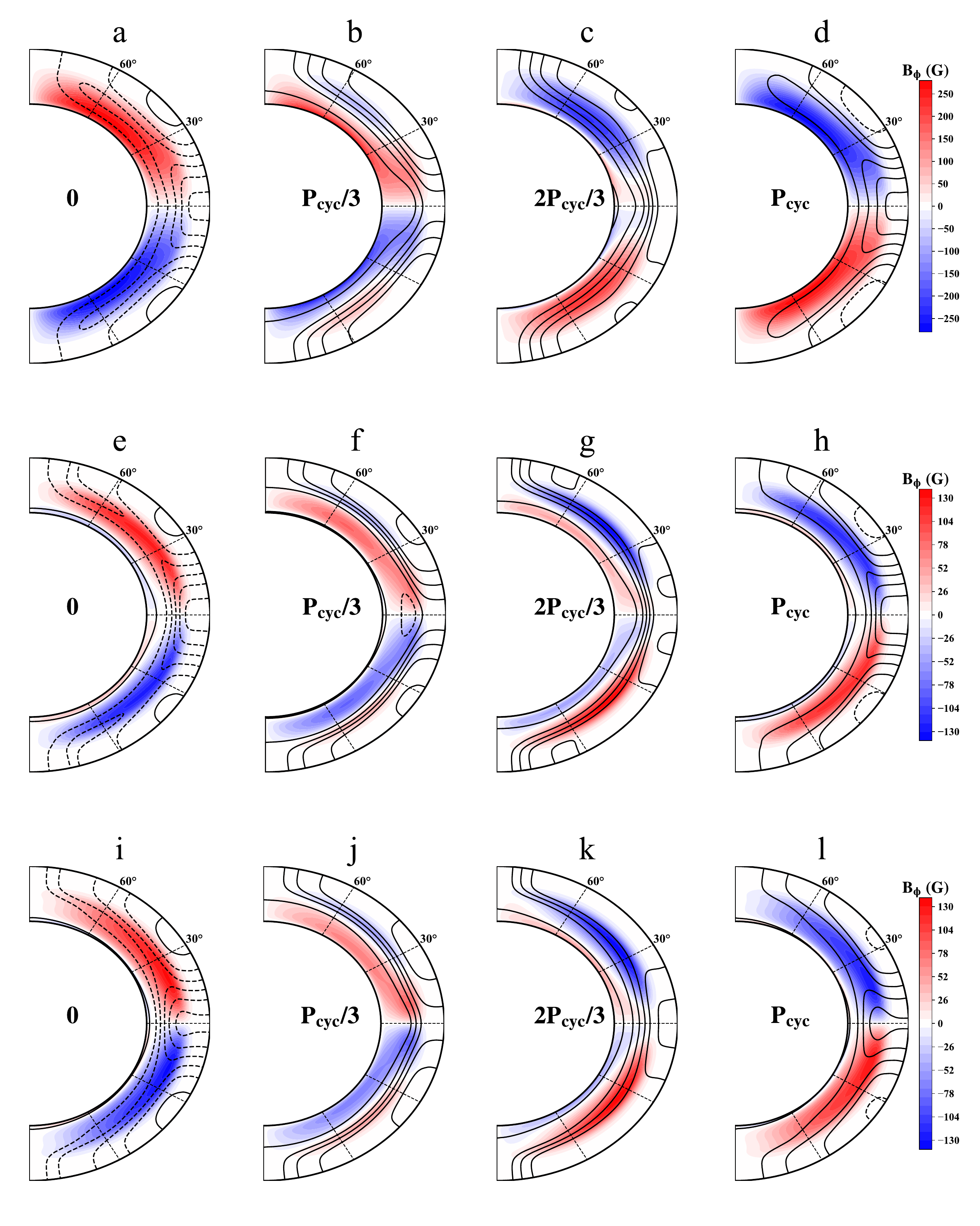}	
	\caption{Snapshots of the toroidal (color shades) and poloidal (contours) field evolution in the meridional cut over one cycle. The top, middle, and bottom rows correspond to the results obtained using the meridional flow profiles MF1, MF2, and MF3, respectively. Each consecutive frame in a row represents a time interval of one-third of the cycle period.}
	\label{Figure6_snap}
\end{figure*}

The stable oscillatory solutions of the distributed-shear BL dynamo model arise from the realistic treatment of the surface field evolution and its effect on the interior. Figures \ref{Figure5_BrThetaT} (a), (b), and (c) present the time-latitude diagrams of the surface radial fields for the three cases with MF1, MF2, and MF3, respectively. All three exhibit features comparable to the observed and simulated magnetic butterfly diagram \citep{Wang2020, Yang2024, Wang2025}, with strong poleward migration branches at high latitudes and weaker equatorward branches at low latitudes. According to \cite{Cameron2012}, the radial outer boundary condition and near surface pumping lead to the dynamo behavior at the surface consistent with SFT models \citep{Jiang2014, Yeates2023}. While the poleward migration branches are also present in the purely latitudinal differential rotation case by \cite{Leighton1969} (see his Fig. 2), these branches are much weaker compared to the low-latitude branch due to the absence of surface meridional flow. The difference between the time-latitude diagrams with and without the meridional flow in SFT simulations can be seen in Figures 7 (c) and (d) of \cite{Baumann2004}. The surface meridional flow and turbulent diffusion work on the poloidal field source and produce the global dipolar field at each cycle minimum. The different concentration of the polar field between Fig. \ref{Figure5_BrThetaT} (a) and Figs. \ref{Figure5_BrThetaT}(b) and (c) arises from the variations in the meridional flow distribution near the poles, as presented in Fig. \ref{Figure1_MF}. Although we have shown that the subsurface meridional flow plays a negligible role in producing the solar-like butterfly diagram, the surface meridional flow is essential for the surface field evolution. This, in turn, affects subsurface toroidal field migration, as will be further demonstrated in Fig. \ref{Figure6_snap}. The role of surface meridional flow will be further explored in the subsequent section. 

Figure \ref{Figure6_snap} presents snapshots of the magnetic field evolution for the 3 meridional flow configurations (upper panels: MF1, middle panels: MF2, lower panels: MF3). Each snapshot corresponds to a time interval of one-third of the cycle period, $P_{cyc}$/3. For all the 3 cases, the interior poloidal fields exhibit a large-scale structure dominated by $l$=1 or $l$=3 and $m$=0 modes, where $l$ and $m$ are the spherical harmonic degree and order, respectively \citep{Luo2024}. In contrast, in the purely latitudinal differential rotation case by \cite{Leighton1969}, higher degree $l$=3 and $l$=5 modes, which correspond to smaller-scale structures, dominate. For FTD models \citep[e.g.,][]{Dikpati1999}, the subsurface meridional flow tends to produce even higher degree modes $l$ (smaller structures). The higher-degree $l$ modes decay more quickly. Therefore, compared to the decay solution supposed to be given by \cite{Leighton1969}, our distributed-shear BL dynamo tends to have a stable oscillatory solution.

Figure \ref{Figure6_snap} also illustrates that two factors contribute to the time and latitude dependent regeneration pattern of the toroidal field. The first factor is the latitude-dependent latitudinal shear, as have illustrated by Fig. \ref{Figure4_LatDepShear}. For a uniform $B_\theta$, the latitudes closer to $\pm55^\circ$ regenerate the toroidal field with a critical amplitude earlier. The second factor arises from the equatorward migration of the surface $B_r$ field, driven by the competition between the equatorward diffusion and the poleward flow at the surface (see each group of panels from left to right) \citep{Petrovay2020, Wang2024}. Due to the source term in Eq. (\ref{eq1:dynamo_A}), the surface $B_r$ of a new cycle, corresponding to the bipolar magnetic region emergence, first occurs around $\pm40^\circ$ latitudes. Then on the surface, leading polarity flux is equatorward transported and the corresponding following polarity flux is poleward transported under the effects of the turbulent diffusion and meridional flow. Subsurface $B_\theta$ is transported inward by pumping and then turbulent diffusion. In the bulk of the convection zone, lower latitudes have the $B_\theta$ of the new cycle later, resulting in a delayed regeneration of the toroidal field at the lower latitudes.

In summary, for the distributed-shear BL dynamo, it is the time and latitude dependent regeneration pattern of the toroidal field that is responsible for the observed solar-like butterfly diagram. The latitude-dependent latitudinal differential rotation and surface field evolution contribute to the regeneration pattern of the toroidal field. During the whole cycle period, interior meridional flow plays a negligible role in producing the equatorward migration of the toroidal magnetic field. The different mechanisms for the equatorward migration of the toroidal magnetic field make the distributed-shear BL dynamo strikingly different from the classical FTD models. 

\subsection{Varied Impacts of Meridional Flow Amplitude on Dynamo Cycle Period} \label{subsec:ImactCyclePeriod}
In the previous subsection, we have shown that the distributed-shear BL dynamo differ significantly from the FTD models in the role of subsurface meridional flow in driving the migration of the toroidal magnetic field. In classical FTD models, the meridional flow is believed to play a crucial role in determining the cycle period: faster meridional flow leads to a shorter cycle period, and vise versa \citep{Dikpati1999, Karak2010, Karak2014a}. In this subsection, we investigate the impact of meridional flow amplitudes across the three configurations on the dynamo cycle period, compare the results with those of FTD models, and analyze the new mechanism that governs the solar cycle period.

As we have claimed in Sect. \ref{sec:model}, for each meridional flow profile, we only adjust the flow amplitude such that the maximum value of surface flow, $u_0$, ranges from 10 to 20 $\text{m s}^{-1}$. The differential rotation profile shown in Fig. \ref{Figure2_DF} (a), which closely matches observational data, is used. Other model parameters including magnetic diffusivity and pumping keep unchanged, as detailed in Sect. \ref{sec:model}. Each simulation, corresponding to a distinct flow profile or amplitude, has a unique critical $\alpha_0$ value, $\alpha_c$, which leads to the growth rate of the dynamo solution approximately zero. 

\begin{figure}[h]
	\centering
	\includegraphics[width=8cm]{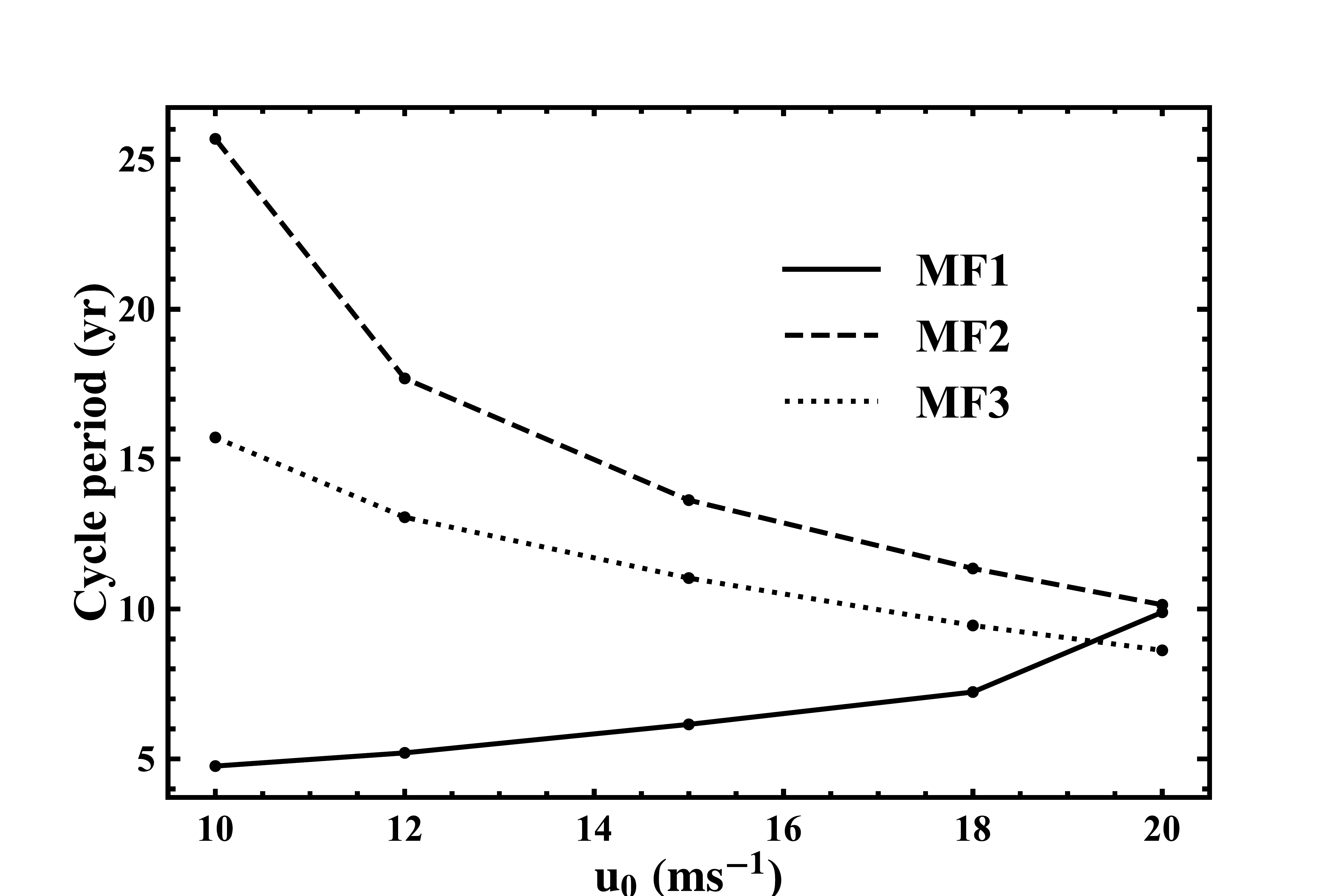}
	\caption{Dependence of the cycle period on the maximum meridional flow speed $u_0$ in distributed-shear BL dynamo models. The solid, dashed, and dotted curves correspond to the results obtained using the meridional flow profiles MF1, MF2, and MF3, respectively. Each simulation is conducted in the linear critical regime, with its own corresponding critical value $\alpha_c$. For each flow profile, we only adjust $u_0$ and other model parameters including magnetic diffusivity and pumping keep unchanged.}
	\label{Figure7}
\end{figure}

The solid line in Fig. \ref{Figure7} shows how the cycle period $P_{cyc}$ of the dynamo solution varies with the maximum flow speed $u_0$ for the MF1 case, which is the configuration used in FTD models. We observe that the cycle period $P_{cyc}$ increases with $u_0$, following the relation $P_{cyc} \propto u_0^{1.1}$ based on a least-squares fit. This highlights another fundamental difference from FTD models, in which case, the power index describing the relation between $P_{cyc}$ and $u_0$ is negative. Specifically, in the circulation-dominated regime, the power index is around -0.9 \citep{Dikpati1999, Yeates2008}, and in the diffusion-dominated regime, it is approximately -0.7 \citep{Chatterjee2004, Karak2010}. The relation between $P_{cyc}$ and $u_0$ has been well understood and applied in previous plethora of FTD models. In the diffusion-dominated regime, the meridional flow’s influence is diminished by diffusion \citep{Jiang2007}. The positive power index observed in our distributed-shear BL dynamo with the deep penetration of the meridional flow, again raises the question: what is the underlying mechanism driving this difference?

The underlying mechanism is primarily attributed to the generation of the surface polar field. For the MF1 configuration, the flow is equatorward in the region between 0.7 - 0.86$R_\odot$, where corresponds to the area of strong latitudinal shear and hence the location of the strong toroidal field, as presented in Figs.\ref{Figure6_snap} a-d. Similar to the behavior in classic FTD models, the toroidal fields are advected equatorward by the flow in the region. The amplitude and latitudinal location of the poloidal source are proportional to the toroidal field, see Eqs.(\ref{eq:S_BL})-(\ref{eq:alpha_f}). A larger $u_0$ results in a more equatorward location of the poloidal source term. As emphasized in \cite{Zhang2022}, the surface $B_r$ evolution in our distributed-shear BL dynamo model, with a radial outer boundary condition and near-surface shear pumping \citep{Cameron2012}, is consistent with results from SFT simulations \citep{Jiang2014, Yeates2023}. According to \cite{Jiang2014b, Petrovay2020}, the lower-latitude emergence of bipolar magnetic regions contributes more to the polar field at cycle minimum with a Gaussian dependence on latitude. Thus, a larger $u_0$ leads to a stronger polar field at cycle minimum. Note that our models are linear, and we normalize the surface $B_r$ to a maximum of 10 G. Therefore, a stronger polar field corresponds to a stronger toroidal flux generated in the bulk of the convection zone, which takes longer time to be canceled by turbulent diffusion. As a result, a faster meridional flow leads to a dynamo solution with a longer cycle period in the distributed-shear BL dynamo model with a deep penetration of the meridional flow.

However, the above relationship between the cycle period $P_{cyc}$ and the maximum flow speed $u_0$ does not hold in the distributed-shear dynamo models when MF2 or MF3 are used. The dashed and dotted curves in Fig. \ref{Figure7} show how $P_{cyc}$ varies with $u_0$ for the MF2 and MF3 cases, respectively. They both present an opposite trend compared to the MF1 case. Here the cycle period decreases as the flow speed increases. The least-squares fits give $P_{cyc} \propto u_0^{-1.4}$ for MF2 and $P_{cyc} \propto u_0^{-0.86}$ for MF3. The behavior contrasts with the MF1 case, suggesting that the phenomenon where a larger $u_0$ results in a more equatorward location of the poloidal source term does not apply to MF2 or MF3. This is true. For MF2, there is no flow between 0.7-0.82$R_\odot$. Although there is an equatorward flow between 0.82-0.9$R_\odot$, the toroidal field in this region is weak (See Figs. \ref{Figure6_snap} e-h). As a result, the effects of the subsurface return flow on the regeneration latitude of the poloidal field at the surface is minimal. A similar explanation applies to the MF3 case. The key to understanding the relationship between $P_{cyc}$ and $u_0$ lies in the effect of the poleward surface flow on the polar field generation and consequently, the toroidal flux generation. As shown in Fig. 7(b) of \cite{Baumann2004}, in the context of the SFT model, the polar field decreases with the increases of the surface flow when the flow exceeds few ms$^{-1}$ for a given surface diffusivity. Accordingly, in the distributed-shear BL dynamo with the MF2 or MF3, a faster meridional flow at the surface weakens the polar field and toroidal flux, leading to a shorter cycle period. This effect also occurs in the MF1 case. But the latitudinal location of the poloidal source has a more pronounced impact due to the Gaussian dependence on latitude \citep{Jiang2014b, Petrovay2020}. As a result, the MF1 case presents an opposite dependence between $P_{cyc}$ and $u_0$. 

Although the distributed-shear BL dynamo models exhibit different power law indices between $P_{cyc}$ and $u_0$ for different meridional flow configurations, the underlying mechanism remains the same. That is the surface flux transport process for the polar field generation, which corresponds to the $\alpha$-effect in the jargon of dynamo theory. In contrast, the surface poloidal field source term plays a minimal role in determining the cycle period in FTD models \citep{Dikpati1999}. The emergence and subsequent evolution of bipolar magnetic region on the surface are central to the BL-type dynamo. In this sense, our distributed-shear BL dynamo models capture the essence of the BL-type dynamo. 

Since the surface flux transport process driving the polar field generation underlies the cycle period of the distributed-shear BL dynamo, the three power law indices between $P_{cyc}$ and $u_0$ are not sufficient to fully describe their relationship. Their relation depends not only on the subsurface flow configuration but also on the specific parameters, such as the diffusivity and the gradient of the flow near the equator at the surface \citep{Petrovay2020,Wang2021} and the diffusivity in the convection zone. We note that switching on or off the near-surface shear layer also affects the results. \cite{Vashishth2024} show a non-monotonic relationship between $P_{cyc}$ and $u_0$, where $P_{cyc}$ decreases initially but then increases with higher flow speeds. Although their model is categorized as an FTD model by them, the surface evolution aligns more closely with SFT behavior due to the radial boundary condition and the near surface radial pumping they adopt. Thus their model can be considered a distributed-shear model in a certain sense. The non-monotonical relationship between $P_{cyc}$ and $u_0$ is one possible form of the distributed-shear BL dynamo. In short, there is no universal power-law relationship between $u_0$ and $P_{cyc}$, as $u_0$ is not the dominant factor of the cycle period, unlike in FTD models. 

\begin{figure}[h]
	\centering
	\includegraphics[width=8cm]{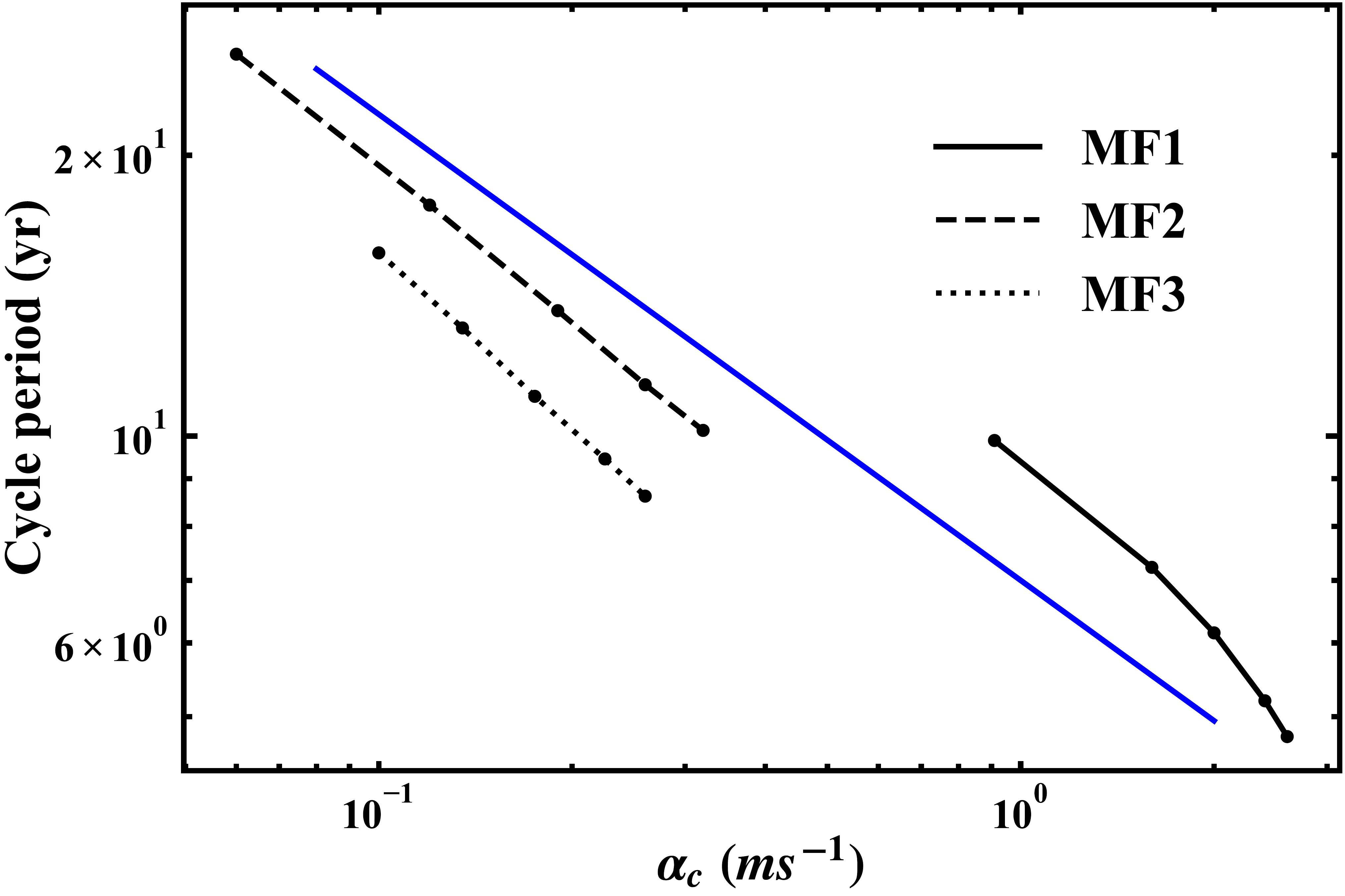}
	\caption{Dependence of the cycle period $P_{cyc}$ on the critical $\alpha_0$ value, $\alpha_c$ in distributed-shear BL dynamo models. The solid, dashed, and dotted curves in black represent the results obtained using the meridional flow profiles MF1, MF2, and MF3, respectively. The blue curve corresponds to $P_{cyc} \propto \alpha_c^{-0.5}$.}
	\label{Figure8}
\end{figure}

We have seen that in the distributed-shear BL dynamo, the $\alpha$-effect, driven by surface polar field generation, plays an important role. This allows the uniform underlying mechanism governing the various relationships between $P_{cyc}$ and $u_0$ to be expressed through the relationship between $P_{cyc}$ and the critical dynamo number, $N_{Dc}$. Given the observationally determined differential rotation and the specified diffusivity, $N_{Dc}$ corresponds to $\alpha_c$. Hence we expect a uniform relation between $P_{cyc}$ and $\alpha_c$, which is illustrated in Fig. \ref{Figure8} for the three meridional flow configurations. In all three cases represented by the three black curves, a decrease in cycle period corresponds to an increase in $\alpha_c$. The blue curve follows $P_{cyc}\propto \alpha_c^{-0.5}$. Notably, all three cases exhibit a similar trend, which is strikingly consistent with the behavior observed in conventional mean-field $\alpha\Omega$ dynamos, as described by the dynamo number, for example Eq. (3.16) in \cite{Yoshimura1975}. This indicates that unlike FTD models, the subsurface flux transport process driven by meridional flow is no longer necessary. This brings the model closer to the conventional $\alpha\Omega$ dynamo.

\section{Conclusions and Discussion} \label{sec:conclustion}
In this paper, we explore the influence of subsurface meridional flow in our distributed-shear BL-type dynamo model \citep{Zhang2022}. Contrary to the FTD framework, we find that subsurface meridional flow plays a minimal role in generating the solar-like butterfly diagram and modulating the cycle period. Remarkably, the characteristic butterfly diagram emerges even in scenarios where the return flow is absent or poleward at the base of the convection zone. The diagram arises from the time- and latitude-dependent regeneration of the toroidal field, driven by latitude-dependent latitudinal differential rotation and surface field evolution. Furthermore, we demonstrate that, unlike in FTD models, the cycle period does not necessarily increase as the flow amplitude decreases. Depending on model parameters, no universal relationship exists between them, as the return flow does not dominate the cycle period. Instead, the cycle period is determined by the surface flux source and transport process responsible for polar field generation, which is central to the BL mechanism and resembles the $\alpha$-effect in mean field dynamos. The butterfly diagram and cycle period highlight the distinct role of surface magnetic field dynamics in distributed-shear BL-type dynamo models.

The distributed-shear BL-type dynamo models benefit from a method that prevents radial diffusion across the surface, ensuring that surface field evolution aligns with the empirically well-constrained SFT models \citep{Wang1989b, Jiang2014, Yeates2023}. The radial outer boundary condition and the introduced near-surface pumping \citep{Cameron2012} have profound consequences for the operation of these models, such that the additional process in contrast to the conventional $\alpha\Omega$ dynamo, that is flux transport below the surface to connect the surface and the tachocline, on longer plays an essential role. As a result, distributed-shear BL-type dynamo models differ strikingly from FTD models. The BL mechanism, which is responsible for the surface field evolution, dominates the cycle period and contributes to the generation of butterfly diagram. In this sense, the BL mechanism, function like the $\alpha$-effect, works in concert with the $\Omega$-effect, making distributed-shear BL-type dynamo models closely align with the original BL dynamo and the conventional $\alpha\Omega$ dynamo.

The method that prevents radial diffusion across the surface also offers an additional advantage for the distributed-shear BL dynamo. Specifically, it allows the magnetic diffusivity in the bulk of the convection zone, $\eta_{cz}$, to be an order of magnitude higher than that in FTD models, where the typical value of $\eta_{cz}$ is on order of $10^{10}$ cm$^{2}$ s$^{-1}$ (See Figure 1 of \cite{Munoz-Jaramillo2011} for illustration). When the near-surface shear layer (i.e., Fig.\ref{Figure2_DF}b), is adopted as the profile of $\Omega$ in Eq. (\ref{eq2:dynamo_B}), stable oscillatory solutions can still be derived even with $\eta_{cz}$ as high as $10^{12}$ cm$^{2}$ s$^{-1}$, which is close to $10^{13}$ cm$^{2}$ s$^{-1}$ estimated from mixing-length theory. The maximum $\eta_{cz}$ that can be sustained by the distributed-shear BL dynamo will be explored in future work.

Not all BL-type dynamo models including near surface radial pumping and a radial outer boundary condition qualify as the distributed-shear BL-type dynamo. For a model to be classified in this category, radial diffusion across the surface should be suppressed, ensuring that the evolution of the surface radial magnetic field aligns with observational constraints \citep{Jiang2014b,Jiang2015}. The toroidal field should be generated in the bulk of the convection zone by the latitudinal shear working on the large-scale field. In the recent BL-type dynamo model by \cite{Cloutier2023}, the adopted parameters fail to satisfy these critical criteria. Specifically, their model adopts a radial pumping speed of $\gamma_{0} \leqslant$ 16.3 m s$^{-1}$ with a penetration depth of $r_p = 0.785 R_\odot$, and surface diffusivity of 3.5 $\times 10^{12}$ cm$^{2}$ s$^{-1}$. These parameters are insufficient to fully suppress radial diffusion across the surface. Moreover, the radial pumping is not confined to the near-surface region, but extends down to $0.785 R_\odot$. As a result, their setup still requires an equatorward return flow at the base of the convection zone to sustain the dynamo process, a feature that distinguishes it from the distributed-shear BL-type dynamo framework. Additionally, we comment that strong near-surface pumping is a method to prevent radial diffusion across the surface. Other approaches could also achieve this effect. One such approach, currently under investigation (Luo et al., in prep.), involves introducing a second-order derivative outer boundary condition to replace the near-surface pumping, which has not been confirmed by observational data.

We note that although the subsurface meridional flow plays a negligible role in the distributed-shear BL-type dynamo, meridional flow remains essential in the models. The poleward surface flow is critical for establishing the Sun's large-scale dipole field. Without this surface component, models tend to produce decaying solutions \citep{Leighton1969, Cameron2017}. Furthermore, the SFT models also have demonstrated that the inclusion of the poleward surface flow is crucial to producing the properties of solar surface field evolution \citep{DeVore1984, Wang1989}. Various methods, including direct Doppler measurements, helioseismology, and feature tracking, have robustly confirmed the existence of surface poleward meridional flows \citep{Jiang2014}. Regardless of the specific profiles of the interior return flows, the distributed-shear BL dynamo can function effectively. The significant role of the surface flow in the distributed-shear BL-type dynamo and its effects on the poleward branch of the torsional oscillation \citep{Zhong2024} warrants further investigation and will be addressed in future work. 

While surface flux transport process plays an essential role in the distributed-shear BL-type dynamo, we caution against labeling such models as a ``flux transport dynamo''. The term traditionally associated with models where subsurface flux transport processes, particularly meridional flow, dominate the production of the butterfly diagram and set the cycle period. The term ``flux transport'' traces its roots to \cite{Leighton1964}, who originally described the dispersal and migration of bipolar magnetic regions due to the random walk of the supergranulation at the surface. Since the surface flux transport is already part of the BL process, emphasizing the term ``flux transport'' within the context of the distributed-shear BL-type dynamo is unnecessary and could create confusion. Even in the future the subsurface meridional flow is confirmed as a single cell with equatorward return flow penetrating to the tachocline, it does not prevent the distributed-shear BL-type dynamo from working properly. 

The solar-like butterfly diagram in the distributed-shear BL dynamo model arises from the latitude dependence of the time required for the toroidal field to reach critical values of buoyant instability. The idea traces back to \cite{Babcock1961}. Over the past six decades, it has been just sporadically referred, such as in \cite{Kopeck1970, Spruit2011}. Recently, \cite{Zhang2022b} demonstrate that a latitude-dependent radial transport mechanism can also produce the butterfly diagram, offering an alternative pathway for achieving time- and latitude-dependent toroidal field regeneration. In addition to the mechanism, other two types of mechanisms are more familiar to the community: the dynamo wave \citep{Parker1955a, Yoshimura1975} and flux transport, typically by meridional flow \citep{Wang1991, Durney1995, Choudhuri1995}. \cite{Guerrero2008,Hazra2016} once proposed the latitudinal component of turbulent pumping as a mechanism for generating the butterfly diagram, which can be categorized as a form of flux transport. \cite{Karak2016} once suggested that near-surface shear layer with downward pumping could account for the butterfly diagram. Actually, this mechanism operates as the dynamo wave. 

We note that the solar-like butterfly diagram is obtained using the linear distributed-shear BL dynamo model in this study. We have also incorporated the nonlinear algebraic $\alpha$-quenching term introduced by \cite{Zhang2022} to assess its impact on the model. Aside from a slight change in the cycle period, the equatorward migration of the toroidal field remains unaffected. More physically motivated forms of nonlinearity will be investigated in future work.

Although the paper demonstrate the success of the distributed-shear BL dynamo model in reproducing solar cycle properties, the dynamo model represents only an initial step towards a realistic BL-type model. The BL mechanism consists of the two key processes: (I) the emergence of toroidal fields in the form of tilted sunspot groups, and (II) their subsequent decay and transport over the surface to generate the large-scale poloidal field. In the current distributed-shear BL-type model, to some sense, Process II has been implemented with a degree of realism. However, Process I, the flux emergence process, remains one of the least understood aspects of solar physics, offering substantial room for refinement. For example, the toroidal flux loss can be modeled in form of double-ring structure \citep{Leighton1969, Durney1995, Munoz-Jaramillo2010} and include latitudinal and tilt quenching nonlinearities \citep{Jiang2020, Karak2020, Talafha2022} in the toroidal flux loss dynamics \citep{Cameron2020}. These aspects are planned for exploration in forthcoming studies.

	\begin{acknowledgements}
We thank the anonymous referee for careful comments, which helped us to improve the manuscript. The research is supported by the National Natural Science Foundation of China (grant Nos. 12425305, 12350004, and 12173005).		
	\end{acknowledgements}

\bibliographystyle{aa} 
\bibliography{references} 

\end{document}